\documentclass[twocolumn, traditabstract]{aa}

\usepackage{natbib}
\usepackage{graphicx}
\usepackage{txfonts}
\usepackage{color}
\usepackage{url}
\usepackage{hyperref}
\usepackage{soul}
\usepackage{xcolor}
\sethlcolor{yellow}
\usepackage[normalem]{ulem}

\newcommand{\footlabel}[2]{%
    \addtocounter{footnote}{1}%
    \footnotetext[\thefootnote]{%
        \addtocounter{footnote}{-1}%
        \refstepcounter{footnote}\label{#1}%
        #2%
    }%
    $^{\ref{#1}}$%
}

\newcommand{\footref}[1]{%
    $^{\ref{#1}}$%
}

\begin{document}

\title{Counting gamma rays in the directions of galaxy clusters.}

\author{D. A. Prokhorov\inst{1}, E. M. Churazov\inst{1,2}}

\offprints{D.A. Prokhorov \email{phdmitry@mpa-garching.mpg.de}}

\institute{Max Planck Institute for Astrophysics, Karl-Schwarzschild-Strasse 1, 85741 Garching, Germany
\and
Space Research Institute (IKI), Profsouznaya 84/32, Moscow 117997, Russia}

\date{Accepted . Received ; Draft printed: \today}

\authorrunning{D. A. Prokhorov, E. M. Churazov}

\titlerunning{Counting gamma rays in the directions of galaxy clusters}

\abstract{Emission of AGNs and neutral pion decay -- are the two most natural
mechanisms, that could make a galaxy cluster be a source of gamma-rays  in
the GeV regime. We revisited this problem by using 52.5-month FERMI-LAT data
above 10 GeV and stacking 55 clusters from the HIFLUGS sample of the X-ray
brightest clusters. The choice of $>$10 GeV photons is optimal from the point
of view of angular resolution, while the sample selection optimizes the
chances of detecting signatures of the neutral pion decay, arising from
hadronic interactions of relativistic protons with an intra-cluster medium,
which scale with the X-ray flux.
In the stacked data we detected a signal for the central 0.25 deg circle at the
level of 4.3 $\sigma$. An evidence for a spatial extent of the signal is
marginal. A subsample of cool-core clusters has higher count rate 1.9$\pm$0.3
per cluster compared to the subsample of non-cool core clusters 1.3$\pm$0.2.
Several independent arguments suggest that the contribution of AGNs to the
observed signal is substantial if not dominant. No strong support for the
large contribution of pion decay was found.
In terms of a limit on the relativistic protons energy density, we got an
upper limit of $\sim$1.5\% relative to the gas thermal energy density, provided
that the spectrum of relativistic protons is hard (s=4.1 in dN/dp=p$^{-s}$).
This estimate assumes that relativistic and thermal components are mixed.
For softer spectra the limits are weaker.}


\maketitle

\section{Introduction}

Galaxy clusters are megaparsec-scale structures that consist of hundreds of galaxies and high temperature, 
$T_{\mathrm{gas}}\simeq 10^{7}-10^{8}$ K, sparse highly ionized plasmas filling the space between galaxies
\citep[for a review, see][]{Sarazin1986}. 
X-ray radiation via bremsstrahlung and via ionic emission lines from intracluster plasmas has permitted to study thermal components of 
the intracluster medium (ICM). 
It has been shown that galaxies and plasmas in a galaxy cluster are gravitationally bound to its dark matter halo.
The constraints on non-thermal components in galaxy cluster cores can be obtained from a comparison of gravitational potential 
profiles derived from X-ray and optical data \citep[e.g., see][]{Churazov2008}.
Detection of diffuse radio emission from many galaxy clusters provided the strong evidence of the presence of relativistic electrons 
with very high energies (with a Lorentz factor of $\gamma\sim 10^{3}$), emitting through the synchrotron mechanism  
in magnetic fields of the ICM \citep[for a review, see][]{Ferrari2008}. 
The interaction of relativistic electrons of such energies with cosmic microwave background (CMB) photons may result in hard 
X-ray emission via the Inverse Compton (IC) process, while the IC emission owing to the interaction of more energetic 
electrons, $\gamma>10^{5}$, with CMB photons may lead to gamma-ray emission \citep[e.g., see][]{Petrosian2008}. 

By analogy with the relativistic particle composition in the Milky Way galaxy, one can assume that relativistic protons are also present in 
galaxy clusters. The collisions of relativistic protons with thermal protons may result in gamma-ray 
emission via the decay of neutral pions produced in these inelastic proton-proton collisions \citep[e.g., see][]{Dermer1986}. 
The candidates as sources of electrons and protons of very high energies in galaxy clusters are active and normal galaxies, 
and structure formation shocks \citep[e.g., see][]{Berezinsky1997}. 
The lifetime of high energy protons, E$>$10 GeV, in galaxy clusters exceeds the Hubble time \citep[e.g., see][]{Volk1996}, 
but the lifetime of very high energy electrons with $\gamma\sim 10^{5}$, which emit gamma-rays via the IC effect, 
in the ICM is several orders of magnitude less than the Hubble time \citep[e.g., see][]{Petrosian2008}. 
Observations of diffuse radio emission from galaxy clusters hosting radio halos indicate that electrons 
with $\gamma\sim10^3$  are continously produced in the ICM and that acceleration mechanisms overtake the electron energy 
losses \citep[e.g.,][]{Brunetti2007}. Note that gamma-rays produced in annihilation of hypothetical dark matter particles, 
such as WIMPs, can potentially contribute to emission from galaxy clusters \citep[e.g., see][]{Fermi2010JCAP}.
  
Clusters of galaxies, which are bright extended X-ray sources, are promising targets for gamma-ray telescopes 
\citep[e.g., see][]{Michelson2010}, but an observational evidence that gamma-rays are emitted by galaxy clusters is still at large. 
Energetic Gamma Ray Experiment Telescope (EGRET) was collecting data on gamma rays (ended in June 2000) ranging from 30 MeV to 30 GeV 
\citep[for a review, see][]{Thompson2008} and its successor, {\it{Fermi}} Large Area Telescope (LAT), 
is collecting the data covering the energy range from 20 MeV to more than 300 GeV \citep[][]{Atwood2009} 
(the science phase of the Fermi mission began on 2008 August 4). 
The analyses of the EGRET data \citep[][]{Reimer2003} and of the first 18 months of {\it{Fermi}}-LAT data 
\citep[][]{Fermi2010ApJ} have allowed to put the flux upper limits on gamma-ray emission from individual galaxy clusters. 
The predicted gamma-ray fluxes from clusters are not strongly below than the observed flux upper limits  \citep[][]{Pinzke2010} 
and this suggests that a stacking analysis should be a powerful tool to search of gamma-rays from galaxy clusters 
\citep[e.g., see][]{Zimmer2011, Dutson2013}. 

At 10 GeV, {\it{Fermi}}-LAT provides an increase in effective area over EGRET on an order of magnitude. 
Note that the AntiCoincidence Detector, ACD, of EGRET was monolithic, which caused problems above 10 GeV: electromagnetic showers 
in the calorimeter produced backsplash which made a signal in the ACD, thus tagging high energy gamma-rays incorrectly as charged 
particles.  To avoid this effect, the ACD for {\it{Fermi}}-LAT is divided into many scintillating tiles\citep[see][]{Moiseev2007} and, 
therefore, {\it{Fermi}}-LAT provides us with a unique opportunity to study incoming gamma-rays with energies above 10 GeV.
At energies below $\approx$10 GeV, the accuracy of the directional reconstruction of photon events detected by {\it{Fermi}}-LAT 
is limited by multiple scattering, whereas above $\approx$10 GeV, multiple scattering is unimportant and the accuracy is limited 
by the ratio of silicon-strip pitch to silicon-layer spacing \citep[see][]{Atwood2007}.

In this paper, we perform a search for gamma-ray emission from a sample of galaxy clusters with modest angular sizes by analyzing
the data from {\it{Fermi}}-LAT.
For this purpose, we stack photons with GeV energies coming from the regions covering clusters of galaxies. 
The theoretical argument that gamma-ray photon spectra of galaxy clusters are harder \citep[e.g., see][]{Miniati2003}  
than that of the galactic and extragalactic diffuse emission \citep[e.g., see][]{Fermi2010PRL, Fermi2012ApJ} provides us with 
lower background at high energies than that those at lower energies and, therefore, provides a higher signal-to-noise ratio at high
energies.  This argument holds for both the mechanisms of gamma-ray production (i.e., via the IC effect and via neutral pion decay).
The fine (0.228 mm) pitch of the strips, applied in {\it{Fermi}}-LAT, gives excellent angular resolution for high energy photons. 
The point spread function of {\it{Fermi}}-LAT strongly depends on photon energy and the 68\% angle containment radius significantly 
decreases with energy (from $\simeq$5$^{\circ}$ at 100 MeV to $\simeq$0.2$^{\circ}$ at 10 GeV, \citet[see][]{Fermi2012ApJS}).  
This allows us to associate more precisely the observed photons above 10 GeV with those that should come from clusters of galaxies, 
and is especially important for galaxy clusters with an angular size smaller than a degree. This precise association permits us 
to perform a temporal analysis for testing the variability of gamma-ray sources in the directions of galaxy clusters. 
Below we demonstrate that the number of high energy photons is sufficient in order to search a statistically significant signal from galaxy 
clusters and that our analysis reveals a tentative excess of gamma rays in the direction of galaxy clusters.
Comparing the positions of galaxy clusters in the sky with those of expected gamma-ray blazar sources from the CGRaBS catalogue 
\citep[see][]{Healey2008}, we exclude eight galaxy clusters possibly associated with CGRaBS gamma-ray candidate sources and put 
the upper limit on relativistic-hadron-to-thermal energy density ratio using a sample of 47 galaxy clusters.     
  
\section{Observation and data reduction}

The {\it{Fermi}}-LAT, the primary scientific instrument on the Fermi Gamma Ray Space Telescope spacecraft, is an imaging 
high-energy gamma-ray  pair conversion telescope covering the energy range from about 20 MeV to more than 300 GeV \citep[][]{Atwood2009}.
Tungsten foils and silicon microstrip detectors are applied to convert incoming photons into electron-positron pairs and to
measure the arrival direction of gamma-rays. A cesium iodide hodoscopic calorimeter is used to provide an energy measurement of the 
electron-positron pairs. 
The Fermi spacecraft travels in an almost circular near-Earth orbit at an altitude of 550 km with a period of about 96 minutes.
The field-of-view of {\it{Fermi}}-LAT covers about 20\% of the sky at any time, and it scans continuously, covering the whole 
sky every three hours (two orbits). By using the silicon strip technique with tungsten convertors (instead of the spark chamber 
techinique), {\it{Fermi}}-LAT achieves a finer precision in the measurement of the directions of incoming gamma-rays than that of EGRET. 
The description of the current instrument performance can be found in \citet[see][]{Fermi2012ApJS}.

We use the first 52.5 months of the {\it{Fermi}}-LAT data, collected between 2008-08-04 and 2012-11-29 and perform the data analysis 
using the Fermi Science Tools v9r27p1 package\footlabel{fn}{\url{http://fermi.gsfc.nasa.gov/ssc/data/analysis/}}. 
``Source'' class events, which are recommended for an analysis of gamma-ray point sources\footref{fn} (including events
converted in both the front and back sections of the LAT tracker) with energies between 10 GeV and 270 GeV are selected. 
Events with zenith angles larger than 100$^{\circ}$ are rejected to minimize contamination from gamma-rays from the Earth limb. 
We removed events  that occur during satellite maneuvers when the LAT rocking angle was larger than 52$^{\circ}$. 
Time intervals when some event has negatively affected the quality of the LAT data are excluded.
We select events in a circular region of interest of 4$^{\circ}$ radius around each galaxy cluster specified below. 

The selection of photon events with E$>$10 GeV for our counting experiment relies on a higher signal-to-noise ratio (S/N) expected
in this energy band compared with that at energies of E$>$200 MeV and E$>1$ GeV. The S/N for a point-like source observed against 
isotropic background is given by 
\begin{equation}
\left(\mathrm{S/N}\right)_{\mathrm{E}}\propto N_{src}(E)\times\sqrt{\frac{A_{\mathrm{eff}}(E)}{I_{\mathrm{bkg}}\times\Sigma}},
\end{equation}
where $N_{\mathrm{src}}(E)$ is a number of incoming photons with energies, $>$E, from the source, $A_{\mathrm{eff}}(E)$ is an effective 
area of the detector as a function of energy, $I_{\mathrm{bkg}}$ is an intensity of background emission, and $\Sigma$ is surface area 
corresponding to the 68\% angle containment radius at energy, E. 
Thus one expects the following improvement of the S/N using E$>$10 GeV instead of E$>$200 MeV
\begin{equation}
\frac{\left(\mathrm{S/N}\right)_{\mathrm{10 GeV}}}{\left(\mathrm{S/N}\right)_{\mathrm{0.2 GeV}}}\approx\left(\frac{10}{0.2}\right)^{-2.1+1}
\sqrt{\frac{0.77}{0.42}\times\left(\frac{10}{0.2}\right)^{(2.4-1)}}\left(\frac{3}{0.25}\right)\approx 3.4,
\end{equation}
and using E$>$10 GeV instead of E$>$1 GeV
\begin{equation}
\frac{\left(\mathrm{S/N}\right)_{\mathrm{10 GeV}}}{\left(\mathrm{S/N}\right)_{\mathrm{1 GeV}}}\approx\left(\frac{10}{1}\right)^{-2.1+1}
\sqrt{\frac{0.77}{0.68}\times\left(\frac{10}{1}\right)^{(2.4-1)}}\left(\frac{0.9}{0.25}\right)\approx 1.5,
\end{equation}
where 0.42, 0.68, and 0.77 m$^2$ are the approximate values of effective area at 0.2, 1, and 10 GeV for normal incidence photons; 
3.0, 0.9, and 0.25$^{\circ}$ are the approximate values of the 68\% angle containment radius at 0.2, 1, and 10 GeV \citep[][]{Fermi2012ApJS}; 
and 2.1 and 2.4 are power-law photon indices for a point source and isotropic background emission, 
respectively. The spectral analysis of gamma-ray observations of galaxy clusters performed in several energy bins 
will be presented in Sect. 4 to clarify the importance of observed photons with energies of several GeV, for a detection of 
gamma-ray sources with hard spectra.
  
As mentioned in Sect.1, X-ray astronomy has provided a perfect tool to detect massive galaxy clusters through imaging and spectroscopy.
ROSAT All-Sky X-ray Survey (RASS; \citet{Trumper1993, Voges1999}) have been performed for the purpose of searching for X-ray 
sources. A highly complete flux limited sample of the X-ray brightest galaxy clusters (HIFLUGCS, the HIghest X-ray FLUx Galaxy 
Cluster Sample) has been published by \citet{Reiprich2002}. Since the gamma-ray luminosities of galaxy clusters are expected to
increase with a cluster mass \citep[e.g., see][]{Pinzke2010} and the X-ray luminosities of galaxy clusters increase with a cluster mass, 
accordingly to the X-ray luminosity--mass relation \citep[e.g., see][]{Hoekstra2011}, we use the HIFLUGCS catalogue to select a sample 
of galaxy clusters for our analysis of the gamma-ray data. Note that galaxy clusters from the HIFLUGCS sample are located at 
high galactic latitudes, $|b|>20^{\circ}$. Therefore, the selection of these clusters for a gamma-ray analysis minimizes contamination 
of a gamma-ray signal from galaxy clusters by diffuse galactic emission which dominates at low galactic latitudes.

The HIFLUGCS catalogue includes fifty seven galaxy clusters with angular sizes (defined by the ratio of the virial radius to the distance) 
smaller than 1$^{\circ}$. These galaxy clusters are targets of our study. The selection of galaxy clusters with high X-ray
fluxes from this catalogue provides us with a suitable sample of galaxy clusters to constrain gamma-ray emission, produced via 
neutral pion decay, using the proportionality of X-ray and gamma-ray fluxes expected in this model \citep[see, e.g.,][]{Ensslin1997}.
The redshifts of these 57 galaxy clusters lie in the interval of 0.016$<$z$<$0.2 (only one cluster lies at z$>$0.1) and their X-ray 
fluxes (in the energy range of 0.1-2.4 keV) are in the interval of (2.0-12.1)$\times10^{-11}$ erg s$^{-1}$ cm$^{-2}$.
This catalogue also contains several groups and clusters of galaxies with large angular sizes exceeding 1$^{\circ}$, 
namely Fornax, Hydra, Coma, NGC4636, Centaurus, and NGC5044. 
Although these six extended objects are ones of the nearest and, therefore, their fluxes are relatively high, we do not include 
them in the present analysis.  This is because their large angular extensions can make problematic to take into account 
a possible population of extragalactic gamma-ray point sources (e.g., blazars), that can contribute to gamma-ray signals 
towards these clusters. Possible inhomogeneities of galactic diffuse emission in the direction of the extended clusters requires 
a supplementary analysis. 
The nearby Fornax, Coma, and Centaurus clusters of galaxies are promising targets for gamma-ray studies in order to constrain the parameters 
of the hadronic cosmic ray model and of some dark matter annihilation models \citep[e.g.,][]{Zimmer2011, Ando2012}.

A gamma-ray signal from clusters of galaxies can be contaminated by various point gamma-ray sources, e.g. blazars, that occasionally
located near galaxy clusters in the sky. To minimize contamination of a signal from galaxy clusters, we explore 
the LAT 2-year Point Source Catalog \citep[the 2FGL catalogue,][]{Nolan2012} searching for point gamma-ray sources 
located near the clusters of galaxies from the HIFLUGCS catalogue and detected by the {\it{Fermi}}-LAT in the first two years 
of the mission. The search shows that there are gamma-ray sources from the 2FGL catalogue near the positions of two galaxy clusters 
(namely, Abell 3376 and Abell 2589). The 2FGL catalogue names of these gamma-ray sources are 2FGL J0602.7-4011 
(which lies at the distance of $\approx$0.26$^{\circ}$ from the center of Abell 3376) and 2FGL J2325.4+1650 
(which lies at the distance of $\approx$0.37$^{\circ}$ from the center of Abell 2589). The gamma-ray source 2FGL J2325.4+1650 
is associated with the radio source NVSS J232538+164641 in \citet[][]{Nolan2012}, while 2FGL J0602.7-4011 is an unidentified 
gamma-ray source.

To check if these two 2FGL sources can possibly be identified with galaxy clusters, we estimate the flux at E$>$200 MeV from each of these 
2FGL gamma-ray sources using their fluxes at E$>$1 GeV and spectral indices listed in the 2FGL catalogue. The calculated fluxes at E$>$200 MeV 
are 6.5$\times10^{-9}$ and 2.4$\times10^{-9}$ ph cm$^{-2}$ s$^{-1}$ for 2FGL J0602.7-4011 and 2FGL J2325.4+1650, respectively, and
are not much lower than the 2$\sigma$ flux upper limit of  4.58$\times10^{-9}$ ph cm$^{-2}$ s$^{-1}$ at E$>$200 MeV for the Coma cluster 
(taken from \citet[][]{Fermi2010ApJ}). 
We use the proportionality relation between X-ray and gamma-ray fluxes (expected for gamma-ray emission via pion decay, see e.g.,
\citet[][]{Ensslin1997}), to test the possible identification of these two 2FGL sources with A3376 and A2589.
The X-ray fluxes for A3376, A2589, and Coma are taken from \citet[][]{Reiprich2002}) and are equal to 4.6$\times10^{-11}$, 4.7$\times10^{-11}$,
and 6.4$\times10^{-10}$ erg cm$^{-2}$ s$^{-1}$, respectively. Taking into account the observed X-ray fluxes and plasma temperatures of
A3376, A2589, and Coma, we found that the expected gamma-ray flux from the Coma cluster is about twenty times greater than those 
expected from A3376 and A2589 when the same relativistic-hadron-to-thermal energy density ratio is assumed.
Since the observed gamma-ray fluxes at E$>$200 MeV from these 2FGL sources are not twenty times lower than the flux upper limit 
for the Coma cluster, we conclude that the upper limit on relativistic-hadron-to-thermal energy density ratio obtained from the analysis 
of the Coma cluster cannot be incorporated with the hypotheses that 1) gamma-rays from 2FGL J0602.7-4011 and 2FGL J2325.4+1650
are produced via neutral pion decay emission in A3376 and A2589 and that 2) the relativistic-hadron-to-thermal energy density ratio
in A3376 and A2589 does not exceed the upper limit derived from the analysis of the Coma cluster.

There are $\approx$1000 BL Lac object type of blazars, FSRQ type of blazars, and other active galactic nuclei 
in the 2FGL catalogue. Assuming that these  objects are isotropically distributed in the sky, the expected number of galaxy  
clusters having one of such 2FGL gamma-ray point sources within the circle of 0.4$^{\circ}$ is $\simeq$0.7 and it is in agreement with 
the fact that there are two galaxy clusters from the sample of 57 clusters in the vicinity of one of 2FGL point gamma-ray sources 
in the sky. Though the possible association of a gamma-ray source with the cluster Abell 3376 is of interest particularly 
in the light of the ring-shaped nonthermal radio-emitting structures discovered in the outskirts of this cluster \citep[][]{Bagchi2006} 
and of the expectation of a strong gamma-ray signal from the outer shock waves of Abell 3376 \citep[e.g.][]{Araudo2008}, we withdraw 
Abell 3376 and Abell 2589 from the sample of the clusters of galaxies selected for the present analysis. 

\begin{figure}
\centering
\includegraphics[angle=0, width=8.0cm]{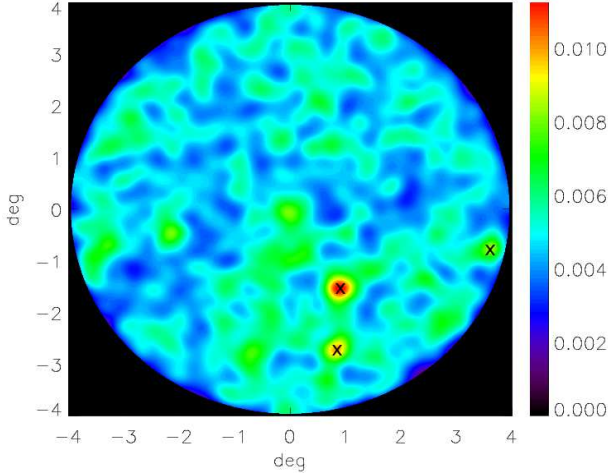}
\caption{Gaussian ($\sigma=0.15^{\circ}$) kernel smoothed count map (before the subtraction of 2FGL sources) 
centered on the positions of galaxy clusters for the energy range 
10 GeV--270 GeV and for a pixel size of 0.005$^{\circ}\times$0.005$^{\circ}$}
\label{F0}
\end{figure}

\begin{figure}
\centering
\includegraphics[angle=0, width=8.0cm]{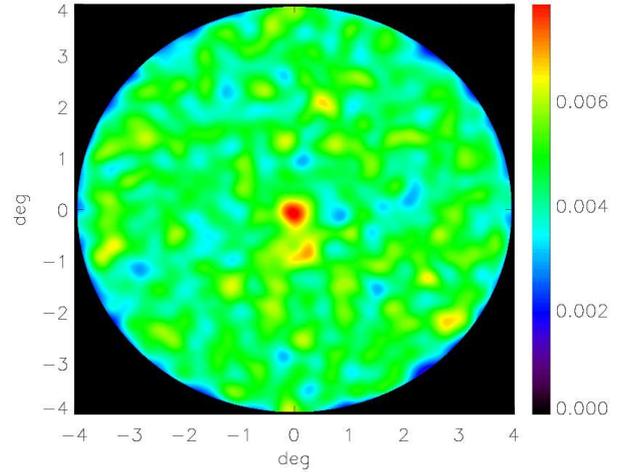}
\caption{Gaussian ($\sigma=0.15^{\circ}$) kernel smoothed count map (after the subtraction of 2FGL sources)
centered on the positions of galaxy clusters for the energy range 
10 GeV--270 GeV and for a pixel size of 0.005$^{\circ}\times$0.005$^{\circ}$}
\label{F1}
\end{figure}

Apart from the point sources located in the close vicinity of galaxy clusters in the sky, there are point sources within the
4$^{\circ}$ regions of interest around each cluster. These point sources should be taken into account in order to estimate diffuse 
background emission. The four brightest of these sources included in the 2FGL catalogue,
namely 2FGLJ0710.5+5908, 2FGLJ1256.1-0547, 2FGLJ1522.1+3144, and 2FGLJ1800.5+7829, are strong at energies higher 10 GeV. 
These four sources are associated with blazars, RGB J0710+591, 3C 279, B2 1520+31, and S5 1803+784, respectively. 
The test-statistics values \citep[][]{Mattox1996} of these four blazars are larger than 100 at E$>$10 GeV accordingly the 2FGL catalogue. 
Since the test-statistics is extensive, one can expect that the test-statistics will be larger 200 for each of these four blazars 
when the first 52.5 months of the {\it{Fermi}}-LAT data are analyzed. 
Thus, e.g., the expected TS value will be equal to 560 for the blazar 2FGLJ1256.1-0547, if we assume that this source is unvariable in time. 
Dividing the expected TS value by the number of the selected regions of interest (i.e., 55), one can expect that this blazar will appear at 
the stacked image as a source with the confidence level of $\simeq$3$\sigma$. Note that the most of gamma-ray sources with high galatic 
latitudes are blazars and that blazars are variable gamma-ray sources in time \citep[e.g.][]{Nolan2012}. 
Flaring activity of blazars can lead to higher TS values (compared with those listed in the 2FGL catalogue) when the 52.5 months of 
the data are analyzed and, therefore, we mask the 2FGL gamma-ray sources to estimate the background emission around the selected galaxy clusters.
The masked 2FGL gamma-ray sources are listed in Table\ref{Table1}. 
In this Table, we show the names of 2FGL sources, the square root of the TS values at E$>$10 GeV (from the 2FGL catalogue), 
and the names of galaxy clusters located within a circle of 4$^{\circ}$ radius around each of the sources. 
To mask these 2FGL sources, we use a circle with a radius of 0.5$^{\circ}$, corresponding to the $\approx$90\% angle containment radius, 
centered on the positions of these sources. 
The stacked count map of the selected galaxy clusters obtained by combining the count maps for each of the galaxy clusters 
from our sample with equal weights and smoothed with a Gaussian kernel  ($\sigma=0.15^{\circ}$)  is shown in Fig. \ref{F0} (and Fig. \ref{F1})
before (and after) the subtraction of 2FGL point sources. Each count map is aligned in galactic coordinates. 
We do not scale the count maps because the virial radii of the selected galaxy clusters are in a narrow range of 0.4$^{\circ}$--0.65$^{\circ}$.
The crosses in Fig. \ref{F0} show the positions of bright 2FGL gamma-ray sources, 2FGLJ0710.5+5908, 2FGLJ1256.1-0547, and 2FGLJ1800.5+7829.
Note that the counts of isotropic background emission that subtracted along with the 2FGL point sources do not strongly 
affect the estimation of isotropic background emission. This is because the total surface area covered by the masked regions is only 
$\approx 1.8$\% of the total surface area covered by the regions of interest around the selected clusters of galaxies.

Figure \ref{F1} demonstrates the excess of the number density of counts towards galaxy clusters (i.e., towards the central region of this 
count map) over the average number density of counts. Studies of spatial, temporal, and spectral properties of the observed tentative 
signal towards galaxy clusters will be presented below. There are three possible interpretations of this central excess:
1) it is due to gamma-ray emission from galaxy clusters; 
2) it is due to a high local concentration of non-2FGL point sources towards galaxy clusters; and
3) it is due to processes occurring in the ICM, but not directly related to galaxy cluster emission.  
The evaluation of statistical significance of this gamma-ray signal towards galaxy clusters and the study of the origin of 
the central count density excess will be present in the Sect. 3.

\section{Results}

In this Section, we evaluate the statistical significance of the observed gamma-ray signal towards the selected clusters of galaxies
and study various interpretations of this signal. 

\subsection{Evaluation of the significance of the signal}

\begin{figure}
\centering
\includegraphics[angle=0, width=9.5cm]{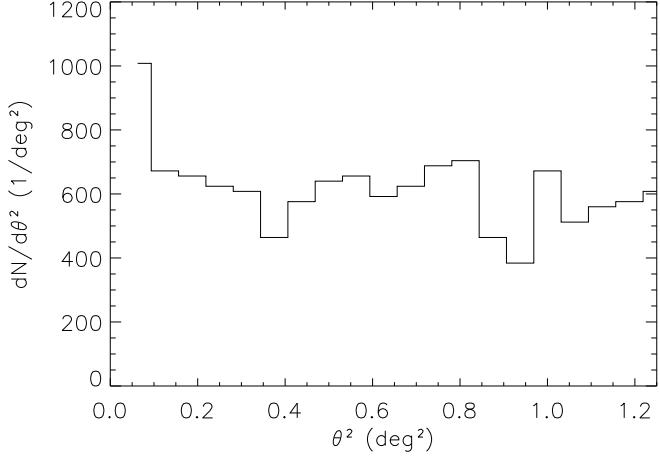}
\caption{Radial distribution of photons in annuli with surface area of $\pi(0.25^{\circ})^2$}
\label{F2}
\end{figure}

To evaluate the statistical significance of the observed gamma-ray signal towards the selected galaxy clusters, 
we bin the count map (its smoothed version is shown in Fig. \ref{F1}), into concentric annuli. 
We choose the surface area of each of the annuli equal to $\pi(0.25^{\circ})^2$. This value of surface area corresponds 
the $68\%$ containment angle at 10 GeV. Thus, the first annulus, which covers the center of the count map, 
should contain $\simeq$68\% of gamma-rays incoming from these galaxy clusters if the signal is strongly centrally concentrated. 
The radius of the k-th annulus is given by R$_{\mathrm{k}}$=$\sqrt{k}\times0.25^{\circ}$. Note that the widths of annuli 
strongly decrease with radius, but this does not affect an estimate of isotropic backround emission.
The numbers of photon events in the first twenty annuli (starting from the center of the count map) are shown in Fig.\ref{F2}. 
These twenty annuli cover the circular region with a radius of $\approx1.1^{\circ}$. The radial distribution of photons, 
shown in Fig. \ref{F2}, confirms the presence of a high count density in the center of the count map. 
The number of counts in the first annulus equals 63 and is significantly higher than the numbers of counts in other radial bins.

The expected number of photons from background emission should be estimated in order to calculate the significance level of
the observed central gamma-ray excess. Since the selected galaxy clusters are at high galactic latitudes ($\mid$b$\mid$ $>$20$^{\circ}$), 
the background emission can thus be considered as isotropic. We calculate the expected number of background photons in the first annulus. 
First, we estimate the expected number of background photons using the nineteen outer annuli. 
The mean value of photons within an annulus is 37.1 and the standard deviation equals 5.2. 
Second, we calculate the expected number of background photons taking the regions of 4$^{\circ}$ radius and excluding 
their most central parts of 0.25$^{\circ}$. By means of the second method, we found that the mean value of background photons 
within an annulus is 36.7 and is in good agreement with that calculated by using the first nineteen annuli.
Therefore, the observed number of counts within the first annulus equal to 63 is significantly higher than the expected number of background
counts, i.e. 36.7. 
The statistical significance level of the count density excess in the first annulus over the expected value equals
$(63-36.7)/\sqrt{36.7}\approx4.3\sigma$. 

We perform the spatial analysis of the central gamma-ray excess to check if the radial distribution of counts agrees with that of 
a point source or of an extended source. Note that the mean virial radius of the selected galaxy clusters corresponds to 
$\approx 0.5^{\circ}$ and, therefore, if the observed excess is produced due to emission via neutral pion decay then this signal should 
be modestly extended and its maximal brightness should be at the center. Assuming that the gamma-ray brightness profiles follow those 
for X-ray brightness derived from a beta model and taking the values of the parameters for a beta model for the selected galaxy clusters 
from the HIFLUGCS catalogue \citep[][]{Reiprich2002}, we found that the 68\% gamma-ray emission from these clusters should be confined 
within the central region with a radius of 0.25$^{\circ}$.
The possible IC gamma-ray emission from the structure formation shocks around galaxy clusters should have an extended profile with maximal 
brightness at shock fronts. However, if gamma-ray emission from galaxy clusters comes from AGNs located in the most central regions 
of galaxy clusters (such as NGC 1275 in center of the Perseus cluster \citep[][]{FermiNGC1275}), the observed radial distribution 
of counts should be described as a point source. 
For studying the spatial extension of the observed signal towards galaxy clusters, we use three models (a point source for emission from 
central AGNs, a disk model for emission produced via neutral pion decay, and a thick ring model for emission produced via the IC radiation 
from structure formation shocks). 
To perform the spatial analysis, we bin the count map into concentric annuli with surface area of $\pi\times(0.25^{\circ})^2/2$. 
The numbers of observed counts in the first sixteen annuli from the center
are 33, 30, 30, 13, 19, 21, 20, 22, 20, 16, 16, 11, 19, 16, 20, and 18. The outer radius of the sixteenth bin is $1^{\circ}$
and corresponds to $\simeq 2$R$_{\mathrm{vir}}$.  

\begin{figure}
\centering
\includegraphics[angle=0, width=9.0cm]{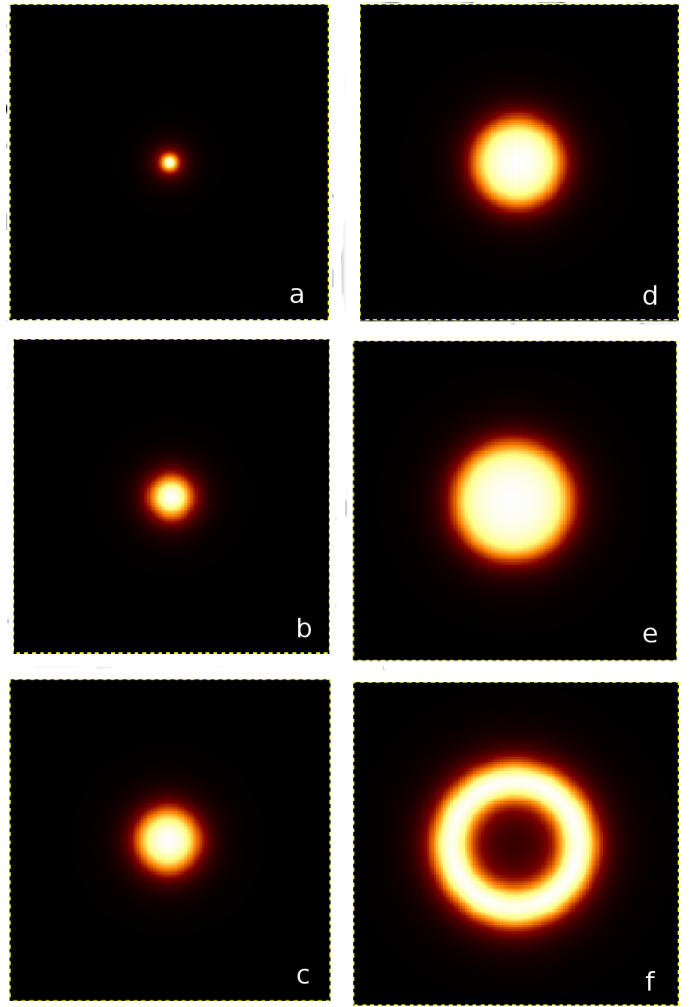}
\caption{Two dimensional templates (with a size of $3\times3$ deg$^{2}$) for the gamma-ray emission from galaxy clusters: 
(a) a point source template; (b-e) templates for emission from a disk of radius 0.15$^{\circ}$, 0.25$^{\circ}$, 
0.375$^{\circ}$, and 0.5$^{\circ}$; (f) a ring-like spatial template}
\label{F3}
\end{figure}

\begin{figure}
\centering
\includegraphics[angle=0, width=9.5cm]{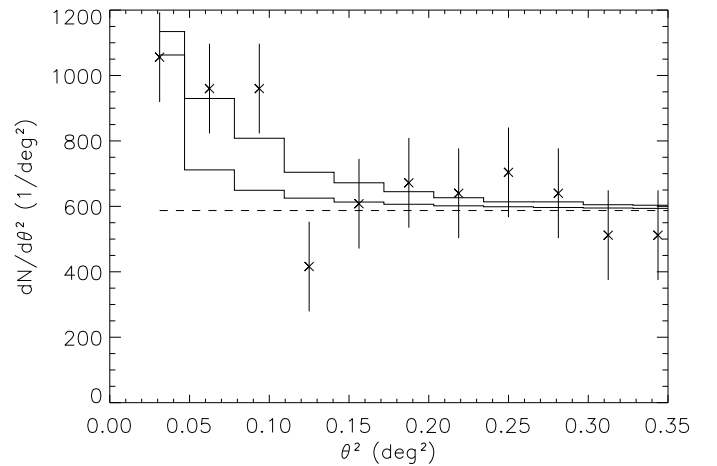}
\caption{The best fits of the point source and disk (R=0.25$^{\circ}$) source models to the data shown by thin and thick solid lines, 
respectively. The surface area of a annular bin is $\pi(0.25^{\circ})^2/2$.}
\label{F4}
\end{figure}

We produce six spatial templates to estimate the significance level of the observed gamma-ray excess.
The first template map describes a point-like source and is obtained by the convolution of a point source model with the PSF.
The following four template maps describe the count distributions for extended disk sources with radii of $0.15^{\circ}$, $0.25^{\circ}$,
$0.375^{\circ}$, and $0.5^{\circ}$ and correspond to the possible neutral pion decay emission from galaxy clusters. 
The sixth template map describes an extended ring-like profile with inner and outer radii of $0.45^{\circ}$ and $0.65^{\circ}$
(correspoding to $\approx$0.9 R$_{\mathrm{vir}}$ and $\approx$1.3 R$_{\mathrm{vir}}$), respectively, and corresponds to the possible 
IC emission from the structure formation shocks around galaxy clusters. All the disk-like and ring-like spatial templates are 
convolved with the PSF. To calculate the PSF, we adopt the power-law spectral distribution for each template with a photon index 
of $-2.1$. The produced templates are shown in Fig. \ref{F3}. To calculate the number of counts, $N_{\mathrm{src}}$, 
observed from the gamma-ray source with one of the spatial profiles, shown in Fig. \ref{F3}, we use the equation
\begin{equation}
N_{\mathrm{src}}=\frac{\sum^{16}_{\mathrm{k=1}} N_{\mathrm{k}} p_{\mathrm{k}}}{\sum^{16}_{\mathrm{k=1}} p^2_{\mathrm{k}}},
\label{eq1}
\end{equation}
where $p_{\mathrm{i}}$ are weights for each of the annuli determined by the spatial profile for a source model and $N_{\mathrm{k}}$ 
is the number of observed counts in the k-th annulus minus the number of counts from isotropic background.
The best fits of the point source and disk source (with a radius of 0.25$^{\circ}$) models to the data are shown in Fig. \ref{F4} 
by thin and thick solid lines. The expected number of counts from isotropic background is shown in Fig. \ref{F4} by a dashed line.
The number of observed counts in each annular bin along with the errorbars is also plotted in this Figure. 
Using the number of counts observed from the source for each of the models, we calculate the significance level of the observed 
gamma-ray excess. The calculated value of the significance level is 4.1$\sigma$ for the point source template.  The values of 
the significance level are 4.6$\sigma$, 4.7$\sigma$, 4.2$\sigma$, and 3.5$\sigma$ for the templates describing gamma-ray emission from disks 
with radii of $0.15^{\circ}$, $0.25^{\circ}$, $0.375^{\circ}$, and $0.5^{\circ}$, respectively. As for the ring-like model, 
we found no evidence for the IC emission around galaxy clusters and that the observed numbers of counts in outer annuli agree with 
that expected from background emission within 1$\sigma$. Thus, the model for gamma-ray emission from structure formation shocks cannot account 
for the observed gamma-ray excess towards the selected galaxy clusters. We list the significance levels for our 
spectral-spatial templates in Table \ref{Table-new}.    

\begin{table}
\centering
\caption{The significance of signal detection}
\begin{tabular}{|c|c|c|}
\hline
Template & Convolution with the PSF & Significance \\ 
\hline
disk (R=0.25$^{\circ}$) & no & 4.3$\sigma$\\
point source & yes & 4.1$\sigma$\\ 
disk (R=0.15$^{\circ}$)& yes & 4.6$\sigma$\\
disk (R=0.25$^{\circ}$) & yes & 4.7$\sigma$\\ 
disk (R=0.375$^{\circ}$) & yes & 4.2$\sigma$\\ 
disk (R=0.5$^{\circ}$) & yes & 3.5$\sigma$\\ 
ring (0.45$^{\circ}<$R$<0.65^{\circ}$) & yes & $<$1$\sigma$\\ 
\hline
\end{tabular}
\label{Table-new}
\end{table} 

We use the sample of 55 galaxy clusters and bin the observed counts in annuli as described above
to calculate the Poisson log-likelihood for each of the spectral-spatial models listed in Table \ref{Table-new}.
Fitting a disk model to the data improves the log-likelihood by $\Delta L\approx$2.5 in comparison to 
the point-source hypothesis. This provides evidence of the spatial extension of the gamma-ray source at a significance 
level of $\sqrt{2\times\Delta L}\approx2.2\sigma$. Note that this significance level is not sufficient to
claim that the gamma-ray source in the directions of galaxy clusters is extended.

We conclude that the significance level of the central gamma-ray excess is greater than 4$\sigma$ for several spatial 
models, such as a point-like source model and a disk-like source model. 

\subsection{The individual contributions of galaxy clusters to the observed gamma-ray signal}

In this Section, we estimate the contribution of each galaxy cluster to the total observed gamma-ray signal towards the selected galaxy clusters
and check the consistency of the observed distribution of the number of photon events in galaxy clusters with a Poisson distribution. 
We also test homogeneity of incoming photons in time to check if the observed gamma-ray signal towards galaxy clusters can be explained by 
time-invariant mechanisms of gamma-ray production, such as those via neutral pion decay or via the IC effect.

\subsubsection{Distribution of the number of photon events in galaxy clusters}

To study the origin of the observed signal towards the selected clusters of galaxies, we perform an analysis of the distribution of 
counts in galaxy clusters. 
Note that the first three annular bins, shown in Fig. \ref{F4}, provide us with the highest signal-to-noise ratio of
$(93-36.7\times1.5)/\sqrt{36.7\times1.5}\approx5.1$.
Thus, for this study we select counts in the directions of the clusters from our sample within a circle of $0.3^{\circ}$ radius. 
Below, we calculate the expected number of galaxy clusters with a given number of counts within a circle of $0.3^{\circ}$ radius 
and compare the expected and observed numbers of galaxy clusters.

The total number of counts in the directions of the galaxy clusters within a circle of $0.3^{\circ}$ radius equals 91. 
Therefore, the average number of counts incoming from the direction of a cluster is given by $91/55\approx1.65$. Assuming the Poisson distribution, 
we calculate the probability to detect a given number of counts towards a galaxy cluster. As expected, the most of the circular regions 
($\simeq75\%$) contains 0, 1, and 2 counts. The expected numbers of clusters with 0--6 counts within a circular region 
of $0.3^{\circ}$ radius are shown in Table \ref{Table2}. If the observed gamma-ray excess towards the selected galaxy clusters is due to
one or two strong gamma-ray sources then the number of circular regions with a high number of counts should be significantly 
higher than that obtained from an estimate based on the Poisson distribution. We calculate the observed numbers of clusters with 
0--6 counts within a circular region of $0.3^{\circ}$ and list the observed numbers of counts in Table \ref{Table2}.   
We find that there is no circular region with a radius of $0.3^{\circ}$ containing more than five counts. 
The difference of the observed number of circular regions with the number of counts, $>2$, and the expected number of such regions
obtained from the Poisson distribution with parameter $\lambda=1.65$ is $\approx -2$. We conclude that the observed number of circular 
regions with a number of counts higher than 2 towards the galaxy clusters is consistent with that obtained from the Poisson distribution with 
parameter $\lambda=1.65$. Therefore, the contribution of one or two possible strong gamma-ray sources to the observed gamma-ray excess cannot 
explain the observed excess.

\begin{figure}
\centering
\includegraphics[angle=0, width=9.5cm]{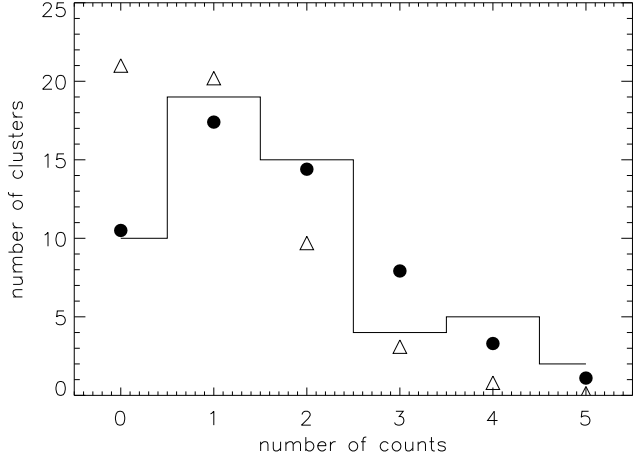}
\caption{The observed number of galaxy clusters with a given number of counts (histogram) and the expected numbers from the Poisson distribution 
with $\lambda=1.65$ (filled circles) and with $\lambda=0.95$ (open triangles)}
\label{F5}
\end{figure}

We also calculate the number of circular regions with a given number of counts that expected from the background emission. 
The average number of counts within a circular region of $0.3^{\circ}$ radius coming from background emission is $52.8/55\approx 0.96$.
The difference of the observed number of circular regions with the number of counts, $>2$, and the number of such regions expected 
from background diffuse emission equals $\approx29$ counts (i.e., $\simeq$55\% of the photons expected from the background emission).
We recalculate the significance of detection of a gamma-ray signal towards the selected galaxy clusters excluding the eleven galaxy clusters 
with the highest observed number of counts within a central circular region of $0.3^{\circ}$ radius and found that the detection significance 
of the excess decreases from $\simeq4.7\sigma$ to 1.5$\sigma$. Therefore, the analysis of these eleven galaxy clusters is important to 
understand the origin of the observed gamma-ray excess. 
In Fig.\ref{F5} we plot the observed and expected distributions of the numbers of galaxy clusters with a given number of counts.

Note that the probability to observe a given number of counts by chance varies from one galaxy cluster to another galaxy cluster 
and depends on the expected number of counts from background around each cluster of galaxy. We calculate the expected
number of background counts for each of the eleven galaxy clusters with the highest observed number of counts. 
To calculate the number of background counts for each circular region with a radius of $4^{\circ}$, we mask 2FGL sources. 
The calculated numbers of background counts are shown in Table \ref{Table3} for these eleven galaxy clusters. 
Using the Poisson statistics, we calculate the probability to obtain the observed number of counts by chance. 
The obtained probabilities are listed in the fourth column in Table \ref{Table3}. Note that the probability to observe 4 or 5 counts in a 
circular region of $0.3^{\circ}$ radius by chance is less than $3\%$. 
 
\begin{table}
\centering
\caption{The expected and observed numbers of galaxy clusters
with a given number of counts}
\begin{tabular}{|c|c|c|}
\hline
Numbers of counts & Expected number & Observed number \\ 
& of clusters & of clusters\\
\hline
0 & 10.5 & 10\\
1 & 17.4 & 19\\
2 & 14.4 & 15\\
3 & 7.9 & 4\\
4 & 3.3 & 5\\ 
5 & 1.1 & 2\\
6 & 0.3 & 0\\ 
\hline
\end{tabular}
\label{Table2}
\end{table} 

\begin{table}
\centering
\caption{Probability to obtain the observed number of counts from background emission by chance}
\begin{tabular}{|c|c|c|c|}
\hline
Cluster name & Observed counts & Background & Probability,$\%$ \\
\hline
A0119 & 3 & 0.73 & 3.1\\
A3112 & 3 & 0.72 & 3.0\\ 
A4038 & 3 & 0.73 & 3.1\\
ZwIII54 & 3 & 1.33 & 10.4\\
A0400 & 4 & 1.05 & 1.8\\ 
2A0335 & 4 & 1.20 & 2.6\\
A1367 & 4 & 0.85 & 0.9\\
MKW4 & 4 & 0.73 & 0.6\\
A2255 & 4 & 1.16 & 2.4\\
A2063 & 5 & 0.86 & 0.2\\
A3581 & 5 & 1.04 & 0.4\\
\hline
\end{tabular}
\label{Table3}
\end{table} 

\subsubsection{Test for homogeneity of incoming photons in time}

Gamma-ray emission from galaxy clusters produced  via neutral pion decay and via the IC emission does not vary with time 
within the time-range of an observation. 
We use this argument to check if the observed gamma rays with E$>$10 GeV towards the selected galaxy clusters are homogeneously 
distributed in time. To perform the test, we choose the galaxy clusters with a high number of observed counts in their directions listed
in Table \ref{Table3}. In Table \ref{Table4} we list the arrival time, given in the Mission Elapse Time (MET) units, and the energy 
for each photons within a circular region of $0.3^{\circ}$ radius centered on A0119, A3112, A4038, A0400, 2A0335, A1367, MKW4, A2255, 
A2063, and A3581. Note that we withdraw ZwIII54 from the sample shown in Table \ref{Table3}, because the probability, $10.4\%$, 
to obtain the observed number of counts from background emission by chance for this galaxy cluster is significantly higher than 
that for other galaxy clusters from Table \ref{Table3}. 
We use the nearest-neighbor method for testing the homogeneity hypothesis and calculate the sum of the distances between each event 
and its nearest neighbor in time. This provides a test statistic that requires no partitioning scheme. Note that the observed number
of counts in a short time interval depends on exposure duration. The exposure duration for successive intervals of time corresponding to 
the precession period of the orbit of the {\it{Fermi}} spacecraft, $\approx55$ days, is roughly constant. 
Since we use the 52.5 months of the data, the observation time contains $\approx28$ time intervals of almost equal exposure.  
Performing the Monte-Carlo simulations for the homogeneous distribution of counts in time, we calculate the number of simulation runs
with the statistic value less than that calculated for each galaxy cluster from Table \ref{Table3}.
The obtained number of simulation runs with the statistic value less than that for A0199 is 12\% of the simulations,
A3112 - 6\%, A4038 - 77\%, A0400 - 54\%, 2A0335 - 48\%, A1367 - 60\%, MKW4 - 57\%, A2255 - 17\%, A2063 - 8\%, and A3581 - 17\%.
We also perform a more powerful test for homogeneity by calculating the probability of several independent simulation runs with a given
number of simulated counts to obtain the test statistic value in each simulation run lower than that is derived from the observation of 
one of the galaxy clusters having the number of counts in its direction the same as that simulated.
The probability that one of three independent Monte-Carlo simulation runs with the homogeneous distribution of 3 photons in time gives 
the statistic value less than that for A0199, and two other simulation runs give the statistic values less than those for 
A3112 and A4038 equals 2.2\%. The probability that one of five independent Monte-Carlo simulation runs with the homogeneous distribution 
of 4 photons in time gives the statistic value less than that for A0400, and four other simulation runs give the statistic values 
less than those for 2A0335, A1367, MKW4, and A2255 is less than 6.3\%. The probability that one of two independent Monte-Carlo 
simulation runs with the homogeneous distribution of 5 photons in time gives the statistic value less than that for A2063, and 
the other simulation run gives the statistic value less than that for A3581 equals 2.0\%. We conclude that the observed gamma rays are
not homogeneously distributed in time and that the observed gamma-ray excess towards the selected galaxy clusters cannot be totally produced 
by the emission via neutral pion decay and by the IC emission. Before starting the search for time variable gamma-ray sources in the
directions of the selected galaxy clusters, we note that different sky-survey profiles have been used on Fermi and, therefore, the on-source 
exposure also varies in time because of the use of different sky-survey profiles. We do not include the exposure variation due to the 
sky-survey profiles in the present analysis and assume that this effect does not affect the result presented in this Section.
In Sect. 3.3, we study the catalogue of AGNs to search of time variable gamma-ray sources towards the selected galaxy clusters.

\subsection{Time variable sources towards galaxy clusters}

Many of the high-latitude 2FGL gamma-ray sources are associated with the bright, flat radio spectrum active galactic nuclei (AGNs)
known as blazars. A uniform all-sky Candidate Gamma-Ray Blazar Survey (CGRaBS) selected primarily by flat radio spectra was 
performed by \citet[][]{Healey2008} to provide a large catalogue of likely gamma-ray AGNs suitable for identification of high-latitude 
gamma-ray sources detected with {\it{Fermi}}-LAT. The CGRaBS catalogue contains 1625 gamma-ray candidate sources with radio 
and X-ray properties similar to those of the EGRET blazars, spread uniformly across the $\mid b\mid >10^{\circ}$ sky. 
81\% of these gamma-ray candidate sources have measured redshifts. Identification of a fraction of gamma-ray sources 
towards the 55 selected galaxy clusters with blazars will allow us to focus on the gamma-rays incoming from galaxy clusters. 
Note that not all detected blazars by {\it{Fermi}}-LAT are included in the CGRaBS catalogue and that $>$1000 sources 
from the CGRaBS catalogue have not yet been detected by {\it{Fermi}}-LAT. 

We perform the search of gamma-ray candidate sources from the CGRaBS catalogue within a circle of $0.6^{\circ}$ (which slightly exceeds
the mean virial radius and the 2 PSF) towards the selected galaxy clusters and found that the gamma-ray candidate sources are present 
towards eight galaxy clusters, namely A0400, A3395s, A3558, A3581, A2029, A2147, A2634, and A2637. 
The distances between each of these galaxy clusters and its nearest gamma-ray source candidate from the CGRaBS
catalogue is 26.6$^{\prime}$ for A0400, 9.9$^{\prime}$ for A3395s, 20.2$^{\prime}$ for A3558, 30.8$^{\prime}$ for A3581,
28.8$^{\prime}$ for A2029, 20.2$^{\prime}$ for A2147, 33.3$^{\prime}$ for A2634, and 31.3$^{\prime}$ for A2657.
We list the redshifts of these galaxy clusters and of the nearest gamma-ray CGRaBS source for each of these clusters, and
the number of counts within a circle of 0.3$^{\circ}$ radius around each of these galaxy clusters in Table \ref{Table5}. 
This Table shows that six of these eight blazars have redshifts larger than 0.35.
Note that two of these galaxy clusters, namely A0400 and A3581, are also present in Table \ref{Table3}. 
The associated CGRaBS gamma-ray candidate sources with A0400 and A3581 are very distant and have redshifts of 1.381 and 2.43, respectively.

\begin{table}
\centering
\caption{List of galaxy clusters possibly contaminated by CGRaBS sources, located at the projected distance $<$0.6$^{\circ}$ from the
cluster centers}
\begin{tabular}{|c|c|c|c|c|}
\hline
Cluster & Number of & Redshift of & Redshift of & Max. energy\\ 
name & counts & a clusters & a blazar & (GeV) \\
\hline
A0400 & 4 & 0.024 & 1.381 & 50.8\\
A3395s & 1 & 0.0498 & 2.051 & 43.5\\ 
A3558 & 1 & 0.048 & 1.326 & 19.7\\
A3581 & 5 & 0.0214 & 2.43 & 16.8\\ 
A2029 & 2 & 0.0767 & 0.084 & 21.4\\
A2147 & 0 & 0.0351 & 0.109 & --\\
A2634 & 1 & 0.0312 & 0.372 & 15.1\\
A2657 & 2 & 0.0404 & 0.667 & 24.4\\
\hline
\end{tabular}
\label{Table5}
\end{table} 

We calculated the number of CGRaBS candidate gamma-ray sources within the first sixteen annuli with surface of 
$\pi (0.6^{\circ})^2$ centered on the 55 selected galaxy clusters and found that the first inner annulus contains 8 CGRaBS sources 
while the mean value of CGRaBS sources within an annulus is 2.8. Thus, the probability of the presence of 8 CGRaBS sources within 
the innerest annulus is only $\approx0.57\%$. We also calculated the number of CGRaBS candidate gamma-ray sources within the 
first sixteen annuli with surface of $\pi(0.5^{\circ})^2$ (the radius of the first annulus corresponds to the mean virial radius 
of the 55 selected clusters) and found each of the first two annuli contains 4 CGRaBS sources. The probability of the presence of 
4 CGRaBS sources within such an annulus is only $\approx1.4\%$. We conclude that the number of CGRaBS gamma-ray candidate sources 
towards the selected galaxy clusters exceeds the expected number within the central annulus with a radius of $0.5^{\circ}$ 
(and $0.6^{\circ}$). One of the possible explanations of the high concentration of CGRaBS sources towards these galaxy clusters 
is a statistical deviation from the expectation.
     
Note that high-energy gamma-rays interact with long-wavelength photons of the extragalactic background light producing 
electron-positron pairs (the Breit-Wheeler process), see, e.g., \citet[][]{Kneiske2002, Franceschini2008, Finke2010}. 
The absoption of gamma-rays from distant sources, such as blazars, via e$^{-}$-e$^{+}$ pair creation leads to decrease the observed 
flux at high energies, E$_{\gamma}>20$ GeV \citep[e.g.,][]{Finke2010}. We use the absoption optical depth of gamma-ray photons as 
a function of observed gamma-ray photon energy, calculated by \citet[][]{Finke2010}, to test the possibility to observe high energy 
photons from the distant CGRaBS sources at z=1.381, 2.051, 1.326, and 2.43, towards A0400, A3395s, A3558, and A3581, respectively. 
We list the maximal observed energy of photons towards these galaxy clusters in Table \ref{Table5}. 
We take the values of absoption optical depth from the tables\footnote{\url{http://www.phy.ohiou.edu/~finke/EBL/}}. 
The absorption optical depth for a gamma-ray with observed energy of 50.8 GeV emitted at z=1.38 is $\approx0.57$, 
for a gamma-ray with observed energy of 43.5 GeV emitted at z=2.05 is $\approx0.77$,  for a gamma-ray with observed energy 
of 19.7 GeV emitted at z=1.33 is $\approx0.022$, and for a gamma-ray with observed energy of 16.8 GeV emitted at z=2.43 is 
$\approx0.1$. We calculated the probability of detection of high energy gamma-rays towards these CGRaBS sources
in the framework of the absorption model of gamma-rays on the extragalactic backround light and found that the detection of the highest 
energy gamma-rays towards the most of these CGRaBS sources is consistent with the expectation from the absorption model. 
However, the probability of detection of a 43.5 GeV photon towards A3395s without observing any photons with energies between
10 GeV and 43.5 GeV towards this cluster is only 0.25\% and, therefore, it is hardly possible that the detected 43.3 GeV photon 
was emitted by a distant blazar.   

\subsection{Likelihood analysis of gamma-ray emission in the directions of A0400 and A3581}

We perform a binned likelihood analysis of gamma-ray emission at the positions of the Abell 400 (A0400) and Abell 3581 (A3581) galaxy 
clusters. These two clusters are selected for an analysis because of their presence both in Table \ref{Table3} of clusters with the largest
number of events at E$>$10 GeV and in Table \ref{Table5} of  clusters possibly contaminated by CGRaBS candidate gamma-ray sources.
The presence of possible gamma-ray sources, namely a binary black hole system 3C 75 in A0400 and a blazar PKS 1404-267 in A3581, 
also provides a good reason for a search of gamma-ray emission from these galaxy clusters. 
The test statistic (TS) is employed to evaluate the significance of the gamma-ray fluxes coming from these two galaxy clusters. 
The TS value is used to assess the goodness of fit and is defined as twice the difference between the log-likelihood functions maximized 
by adjusting all parameters of the model, with and without the source, and under the assumption of a precise knowledge of the galactic 
and extragalactic diffuse emission. We use the publicly available tool ({\it{gtlike}}), released by the {\it{Fermi}}-LAT collaboration,
to perform a binned likelihood analysis. 
For each analyzed source, we select events with energy E$>$100 MeV in a circular region of interest of 20$^{\circ}$  radius. 
We use the first 52.5 months of the {\it{Fermi}}-LAT data and the good time intervals such that the region of interest does not go below 
the Earth Limb. We use the spectral-spatial templates, {\it{gal$\_$2yearp7v6$\_$v0.fits}} and {\it{iso$\_$p7v6source.txt}}, 
for the galactic and extragalactic diffuse emission, respectively.  
To produce a spectral-spatial model for each galaxy cluster, we fix their positions at the localization of 3C 75 for A0400 
and of PKS 1404-267 in A3581 and include the gamma-ray sources from the 2FGL catalogue. The spectra of 2FGL sources are taken 
from the 2FGL catalogue. The spectral parameters of the most of 2FGL sources are fixed at the 2FGL values, while the parameters of 
the strongest sources in the region of interest are left free in the likelihood fit. We model the emission from these two galaxy 
clusters under the assumption of a point source with a power-law spectral distribution. The likelihood ratio test shows that the TS 
values for A0400 and A3581 equal 41.7 and 5.3, respectively. The approximate value of the significance of detection is given by
$\sqrt{TS}\sigma$ and, therefore, the significance of detection is 6.4$\sigma$ and 2.3$\sigma$ for A0400 and A3581, respectively.
Note that the possible explanation of a higher significance value for A400 is that diffuse emission is strongly inhomogeneous on a scale
of $\simeq5^{\circ}$ towards this galaxy cluster. The performed likelihood analysis shows that the gamma-ray 
sources towards these galaxy clusters will be promising targets for an analysis in the future.

\subsection{Cool-core and non-cool core galaxy clusters}

Clusters of galaxies with dense gaseous cores have central cooling times much shorter than a Hubble time. 
XMM-Newton observations of cool-core galaxy clusters have shown that the spectral features predicted by the cooling flow models 
are absent in the X-ray spectra \citep[][]{Peterson2001, Peterson2003}. 
The likely source of energy, compensating gas cooling losses, comes from central AGN \citep[e.g.,][]{Churazov2000, David2001, Bohringer2002}. 
The presence of dense gaseous cores in cool-core clusters can lead to the increase of gamma-ray emission due to neutral pion decay produced 
in the collisions of relativisitic protons with protons of the ICM, while the central AGN itself can also be a source of 
gamma-ray emission. Non-cool core clusters in their turn often show evidence of a merger and, therefore, CR hadrons and electrons 
accelerated in merger shocks can lead to gamma-ray emission via the neutral pion decay or IC effect. In this Sect., we perform an analysis of 
a sample of cool-core galaxy clusters and of a sample of non-cool core clusters to study gamma-ray emission of these two distinct types 
of galaxy clusters.

We divide 55 selected galaxy clusters into the two samples using the definition from \citet[][]{Chen2007}. If the ratio of the mass 
deposition rate, $\dot{M}$, to the cluster mass, $M_{500}$, for a galaxy cluster exceeds $10^{-13}$ yr$^{-1}$ then the galaxy cluster belongs 
to the sample of cool-core clusters. The second sample of galaxy clusters consists of galaxy clusters with the ratio, 
$\dot{M}/M_{500}$, smaller than $10^{-13}$ yr$^{-1}$. We take the values of 
$\dot{M}/M_{500}$ and $M_{500}$ from Tables 1 and 4 of \citet[][]{Chen2007}. There are two galaxy clusters, A1651 and A2063, with 
the calculated ratios of $\dot{M}/M_{500}$ close to $10^{-13}$ yr$^{-1}$. 
We include these two clusters in the sample of cool-core clusters.
The sample of cool-core clusters consists of 27 clusters and the sample of non-cool core clusters consists of 28 clusters.

Using the spectral-spatial templates presented above, we calculate the significance of detection of a gamma-ray signal from both the 
samples. The method used to evaluate the statistical significance is analogous to that applied in Sect. 3.1. The
evaluated values of the significance of detection are shown in Table \ref{Table-cool}. We found that the significance of the gamma-ray
signal from the sample of cool-core clusters, $\simeq 5\sigma$, is significantly higher than that from the sample of non-cool core 
clusters $\simeq 2\sigma$. Furthermore, we conclude that the observed excess of gamma-ray emission in the direction of the 55 selected clusters
is mainly owing to gamma rays coming from the cooling core clusters.

Assuming that the numbers of observed photons in both the samples of galaxy clusters can be modeled by a Poisson distribution, 
we check if the two Poisson samples have the same mean. To perform this analysis, we use the test-statistics, 
$TS=(O_{1}-E_{1})^2/E_{1}+(O_{2}-E_{2})^2/E_{2}$, where $O_{1,2}$ are the observed number of photons within a circle of 0.3$^{\circ}$ centered
on galaxy clusters from each sample and $E_{1,2}$ is the number of photons expected under the assumption that the two Poisson samples 
have the same mean. We found that the probability that both the samples have the same mean is 6.3\%. We conclude that the contribution of 
cooling-core clusters to the gamma-ray emission from the 55 selected clusters is dominant over that of non-cooling core clusters. 
Therefore, the gamma-ray properties of the cooling core clusters significantly differs from those of the non-cooling core clusters.

\begin{table}
\centering
\caption{The significance of signal detection for the samples of cooling core and non-cooling core galaxy clusters}
\begin{tabular}{|c|c|c|}
\hline
Template & cooling core clusters & non-cooling core clusters\\ 
\hline
point source & 4.3$\sigma$ & 1.8$\sigma$\\ 
disk (R=0.15$^{\circ}$)& 4.9$\sigma$ & 2.0$\sigma$\\
disk (R=0.25$^{\circ}$) & 5.0$\sigma$ & 2.0$\sigma$\\ 
disk (R=0.375$^{\circ}$) & 4.6$\sigma$ & 1.8$\sigma$\\ 
disk (R=0.5$^{\circ}$) & 3.5$\sigma$ & 1.9$\sigma$\\ 
\hline
\end{tabular}
\label{Table-cool}
\end{table}  

Note that our selection of 55 galaxy clusters is based on the flux limited cluster sample from the HIFGLUGS catalogue.
Therefore, the expected proportionality between X-ray and neutral pion decay induced gamma-ray fluxes \citep[see][]{Ensslin1997} leads to 
a testable prediction about higher gamma-ray fluxes from brighter X-ray galaxy clusters. Taking this prediction into account, 
we perform a likelihood analysis weighting the observed number of gamma rays from each galaxy cluster with its expected gamma-ray flux.
We calculate the Poisson loglikelihood for each of the spectral-spatial models using two weighting techniques: 
1) combining the observed count maps with equal weights (as was done in Sect. 3.1) 
and 2) weighting the observed number of counts with expected gamma-ray fluxes. 
We compute the difference in the maximized values of the loglikelihood (L) functions for both weighting techniques and
for each spectral-spatial model and found that the maximal likelihood for the ``equal weights'' model is larger than that for 
the hadronic model of gamma-emission. 
We use the Akaike information criterion to measure the relative quality of the models supporting these two weighting techniques and 
found that the hadronic model of gamma-emission is $\exp(\mathrm{L_{2}}-\mathrm{L_{1}})\simeq 5\times10^{-3}$ times as probable as 
the ``equal weights'' model to minimize the information loss. 
The significance of detection of a signal in the direction of the 55 selected galaxy clusters is in the range of 2.1-2.7 $\sigma$ 
(depending on the spectral-spatial models) when the weighting 
technique presuming the expected gamma-ray fluxes from the hadronic model is used. 
Therefore, there is no convincing evidence that observed signal is due to neutral pion decay, at least in the simplest version
of the model presuming $F_{\gamma}\propto F_{X}$.

\subsection{Upper limit on relativistic-hadron-to-thermal energy density ratio}

We reevaluate the significance of a gamma-ray signal from clusters of galaxies withdrawing 8 galaxy clusters possibly contaminated 
by CGRaBS gamma-ray candidate sources (see Sect. 3.3) from our sample. Using the method described in Sect. 3.1 and the templates
for gamma-ray emission produced above, we found that the values of the significance level are 3.3$\sigma$, 3.6$\sigma$, 3.2$\sigma$, 
and 2.3$\sigma$ for the disk templates with radii of 0.15$^{\circ}$, 0.25$^{\circ}$, 0.375$^{\circ}$, and 0.5$^{\circ}$, 
respectively. The significance level for the point source template is 2.7$\sigma$. For the ring-like model, we found no evidence 
for an excess over background. Note that the mean number of counts within a circle with a radius of 0.3$^{\circ}$
centered on the galaxy clusters possibly associated with CGRaBS sources is $2.0\pm0.5$, while the corresponding average number calculated for 
the 47 selected galaxy clusters (excluding those that have the possible association with CGRaBS sources) is $1.6\pm0.18$. 
For comparison, the mean background contribution to the same region is $\simeq$0.96 per cluster.
A slightly lower flux in the reduced sample is yet another reason (apart from smaller sample size) why the detection significance 
is lower in the reduced sample. 

The energy density stored in hadronic cosmic rays (CR) relative to the thermal energy density of the ICM gas can be constrained by using
the observed gamma-ray and X-ray fluxes from galaxy clusters \citep[][]{Ensslin1997, Fermi2010ApJ}. Using the Fermi Science 
Tools\footref{fn}, we calculate the exposure weighted by the power-law with a photon index of -2.1 for each of the 47 galaxy clusters
to convert count numbers to gamma-ray fluxes at E$>$10 GeV. We take the X-ray fluxes from these galaxy clusters and their 
plasma temperatures from the catalogues published by \citet[][]{David1993, Reiprich2002}. We take the production rate of gamma-rays 
from \citet[][]{Drury1994} for a spectral index, $s$=4.1, of the parent CR hadron distribution $f(p)\propto p^{-s}$, where
$p$ is momentum, and extrapolate the production rate at E$>$1 TeV to E$>$10 GeV. 
Note that the production rate of gamma-rays is defined as the gamma-ray emissivity normalized to the CR hadron energy density.
Using Eq.(10) from \citet[][]{Ensslin1997} and the calculated production rate of gamma-rays at E$>$10 GeV, we obtain the relation 
between the observed fluxes of gamma-rays, $F_{\gamma}(>\mathrm{10GeV})$, and X-rays, $F_{\mathrm{X}}$, that is given by
\begin{equation}
\frac{F_{\gamma}(>10\mathrm{GeV})}{F_{\mathrm{X}}/\mathrm{erg}}\approx 1.33\left(\frac{k_{\mathrm{B}} T}{\mathrm{keV}}\right)^{1/2} 
\alpha,
\label{eq2}
\end{equation}
where $\alpha$ is the scaling ratio between the thermal, $\epsilon_{\mathrm{th}}$, and CR hadron energy, 
$\epsilon_{\mathrm{CR}}$, densities, $\epsilon_{\mathrm{CR}}=\alpha\times\epsilon_{\mathrm{th}}$.
Note that the gamma production rate may be taken to be proportional to the number density of relativistic hadrons
\citep[][]{Drury1994} 
and that the energy density of relativistic particles only slightly depends on the low energy bound for hard power-law hadron spectra with s=4.1.
It is also noteworthy that soft power-law hadron spectra, e.g. s=-4.7, produce softer gamma-ray spectra and that the value
of CR hadron energy density derived by means of this method depends on the assumed power-law hadron spectra. For softer hadron spectra, the
factor of 1.33 in Eq. \ref{eq2} will be lower (e.g., $\simeq$0.08 for s=4.7) and the contribution
of hadrons of energy below about 1 GeV per nucleon, which do not generate observable gamma rays, 
to the total energy density will be dominant.

To estimate the scaling ratio, $\alpha$, from the observed gamma-ray fluxes in the directions of galaxy clusters,
we rewrite Eq. \ref{eq2} as 
\begin{equation}
\alpha\approx\frac{1}{1.33}\frac{F_{\gamma}(>10\mathrm{GeV})}{F_{\mathrm{X}}/\mathrm{erg}}\left(\frac{\mathrm{keV}}{k_{\mathrm{B}} T}\right)^{1/2}. 
\label{eq3}
\end{equation}
Putting the parameters into Eq.\ref{eq3}, we calculate the scaling ratio $\alpha$ for each of the selected galaxy clusters.
Assuming that the value of the scaling ratio, $\alpha$, is the same in these 47 galaxy cluster, 
we calculate the weighted mean, $\alpha_{\mu}$, and variance of mean, $\sigma^2_{\mu}$, as 
$\alpha_{\mu}=\sum_{\mathrm{k}} (\alpha_{\mathrm{k}}/\sigma^2_{\mathrm{k}})/\sum_{\mathrm{k}} (1/\sigma^2_{\mathrm{k}})$ and 
$\sigma^2_{\mu}=1/\sum_{\mathrm{k}} (1/\sigma^2_{\mathrm{k}})$,
where $\alpha_{\mathrm{k}}$ are the values of the scale ratio for individual galaxy clusters and $\sigma_{\mathrm{k}}$ are the uncertainties 
of $\alpha_{\mathrm{k}}$ (dominated by the uncertainties in $F_{\gamma}$). 
Calculating the weighted mean of the scaling ratio and the variance of the mean, we found that the value of the relativistic-hadron-to-thermal 
energy density ratio equals $\approx$1.5\%. 
The assumption that this gamma-ray signal is caused by sources variable in time (AGNs) allows us to put an upper limit on the
relativistic-hadron-to-thermal energy density ratio equal 1.5\%. The possible relativistic hadron component with energy density of 1.5\% 
of that of the thermal plasma does not violate the constraints obtained by an alternative method by \citet[][]{Churazov2008} and 
its presence should be tested by a combined analysis of the {\it{Fermi}}-LAT observations towards the extended galaxy clusters with 
high X-ray fluxes, such as Virgo, Perseus, Fornax, Hydra, Coma, and Centaurus. An analysis of these extended galaxy clusters is beyond the scope
of our paper and requires a more sophisticated technique to deal with the spatial extention of these nearby X-ray brightest clusters of galaxies.    

We also study a gamma-ray flux-cluster-mass scaling relation for CR induced emission and compute a prefactor, $\beta$, 
in the relation proposed by \citep[][]{Pinzke2011} 
\begin{equation}
F_{\gamma}(>10\mathrm{GeV})=\beta\times10^{43}\left(\frac{1}{4\pi D^2_{L}}\right) \left(\frac{M_{200}}{10^{15} M_{\odot}}\right)^{1.34} 
\mathrm{ph}\cdot\mathrm{s}^{-1}\mathrm{cm}^{-2},
\label{eq4}
\end{equation}
where the gamma-ray flux is dominated by photons produced via the decaying neutral pions and also includes the IC gamma-ray contribution 
from shock-accelerated primary electrons as well as secondary electrons created in relativistic hadron-proton interactions. 
The theoretical estimate for the prefactor, $\beta$, equals 1.4 \citep[][]{Pinzke2011}. By analogy with the method used in 
calculation of the energy density stored in hadronic cosmic rays relative to the thermal energy density, we rewrite Eq.\ref{eq4} as
\begin{equation}
\beta=\frac{F_{\gamma}(>10\mathrm{GeV}) 4\pi D^2_{L}}{10^{43}} \left(\frac{10^{15} M_{\odot}}{M_{200}}\right)^{1.34} 
\label{eq5}
\end{equation}
and calculate the value of $\beta$ for each galaxy cluster, the weighted mean $\beta_{\mu}$, and variance of mean. We found that
the value of the prefactor, $\beta$, equals 3.5$\pm$1.0 and is slightly higher than that predicted by \citet[][]{Pinzke2011}.
Given plausible contribution to the gamma-ray flux from central AGNs, we conclude that observed $\beta\lesssim3.5$.

\section{Spectral analysis in four energy bands}

In the previous Sects., we presented the analysis of gamma-ray emission towards 55 galaxy clusters at energies greater 10 GeV.
The tentative gamma-ray excess towards these galaxy clusters at E$>$10 GeV was detected. To validate the presence of this
gamma-ray excess, below we perform an analysis of gamma-ray emission towards these galaxy clusters in several energy bands. 
We choose four energy bands for our spectral analysis, which are 1924-3333 MeV, 3333-5773 MeV, 5773-10000 MeV, and $>$10 GeV.
Low boundaries of these 4 energy bands are logarithmically spaced and the low energy boundary of the first band is chosen taking 
into account the high spatial resolution at E$\gtrsim$2 GeV sufficient to avoid contamination of a signal from galaxy clusters by 
gamma rays incoming from 2FGL sources.

We use the point and disk source models, described in Sect. 3.1, to calculate the number of observed photons
towards galaxy clusters in the four energy bands  1924-3333 MeV, 3333-5773 MeV, 5773-10000 MeV, and $>$10 GeV. 
To perform the spatial analysis, we bin the count maps into concentric annuli with surface area of $\pi\times\mathrm{PSF}^2/2$.
The evaluated numbers of photons from the source in the directions of galaxy clusters in each energy band are shown in Table \ref{Table6}. 
The energy bands are enumerated in increasing order in this table. To calculate the PSF in each energy band, 
we adopted the power-law spectral distribution for each template with a photon index of 2.1. Table \ref{Table6} shows that 
the number of photons in the third energy band, i.e. 5773-10000 MeV, is greater than zero at a statistical level of 
$\approx 2.5\sigma$. Thus, the third and fourth energy bands provide an evidence for the possible gamma-ray signal from galaxy clusters.

We estimate the significance of a gamma-ray signal towards galaxy clusters by use of the likelihood ratio method \citep[e.g., see][]{Li1983}. 
In the present problem there is one unknown parameter - the expectation of the number of source photons. Note that the expectation
of the number of background photons is derived from outer spatial bins where the number of photons coming from the source is negligible. 
The statistical hypothesis tested here is no extra source exists and all observed photons are due to 
background. If the errors are Gaussian, then the likelihood is given by $\mathcal{L} =\exp(-\chi^2/2)$, so that minimizing $\chi^2$ 
is equivalent to maximizing $\mathcal{L}$. The correctness of the estimation procedure follows from the fact that variable  
$-2\ln(\mathcal{L}/\mathcal{L}_{\mathrm{max}})$
is distributed like $\chi^2$ with one degree of freedom if there is one parameter of the problem \citep[see][chap. 13.7]{Wilks1962}.
We re-evaluate the significance of the gamma-ray excess at energies, $>$10 GeV, and confirm the results presented in
Table \ref{Table-new} of Sect. 3.1 which obtained by means of an alternative method. We perform a combined likelihood analysis 
in the four energy bins. We found that the value of the significance level is 4.8$\sigma$ for the point source template and 
that the values of the significance level are 5.0$\sigma$, 5.1$\sigma$, 4.8$\sigma$, and 4.6$\sigma$ for the templates
describing gamma-ray emission from disks with radii of 0.15$^{\circ}$, 0.25$^{\circ}$, 0.375$^{\circ}$, and 0.5$^{\circ}$, respectively.
Thus, the inclusion of the three low energy bands to the analysis increases the significance of detection of a tentative 
gamma-ray excess in the directions of galaxy clusters.

\begin{table}\scriptsize
\centering
\caption{The evaluated number of counts from the source in the direction of galaxy clusters in four energy bands}
\begin{tabular}{|c|c|c|c|c|}
\hline
Template & N$_\mathrm{counts}$, band1 & N$_\mathrm{counts}$, band2 & N$_\mathrm{counts}$, band3 & N$_\mathrm{counts}$, band4 \\ 
\hline
point source & 37.6$\pm$33.6 & 15.0$\pm$16.8 & 20.3$\pm$7.8 & 27.8$\pm$6.7\\ 
disk (R=0.15$^{\circ}$)& 40.4$\pm$34.7 & 17.4$\pm$18.1 & 22.0$\pm$8.4 & 30.9$\pm$8.5\\
disk (R=0.25$^{\circ}$) & 44.3$\pm$36.1 & 20.7$\pm$19.8 & 24.3$\pm$11.0 & 46.0$\pm$9.7\\ 
disk (R=0.375$^{\circ}$) & 53.7$\pm$39.3 & 28.3$\pm$19.8 & 28.2$\pm$11.0 & 51.1$\pm$12.1\\ 
disk (R=0.5$^{\circ}$) & 68.5$\pm$43.8 & 38.4$\pm$26.7 & 32.2$\pm$13.0 & 49.5$\pm$14.2\\ 
\hline
\end{tabular}
\label{Table6}
\end{table}

\section{Conclusions}

Clusters of galaxies are promising targets for gamma-ray telescopes. Gamma-rays can be emitted from the ICM via the decay of neutral pions 
produced in the inelastic proton-proton collisions or via the IC radiation owing to the interaction of very energetic electrons with CMB photons. 
Apart from these two radiative processes, gamma-rays can also be emitted by the central active galactic nuclei in galaxy clusters. 
In this paper, we searched for 10 GeV emission from galaxy clusters using the first 52.5 months of the {\it{Fermi}}-LAT data. We performed 
a combined analysis of the 55 galaxy clusters selected from the HIFLUCGS catalogue such that the virial radius of
each selected galaxy cluster does not exceed 1$^{\circ}$. We selected the photon events with energies above 10 GeV, incoming from the 
circular regions with a radius of 4$^{\circ}$ and centered on these 55 galaxy clusters. The selection of this energy interval is motivated by 
the high angular resolution of {\it{Fermi}}-LAT at these energies that permits us to associate more precisely the observed gamma-rays 
with those that should come from galaxy clusters.

To evaluate the significance of a  gamma-ray signal in the direction of the 55 galaxy clusters, we produced six spectral-spatial templates 
(see Fig. \ref{F3}) and fitted the models based on these templates to the observed data. The six models include a point source model, 
four disk models with radii of 0.15$^{\circ}$, 0.25$^{\circ}$, 0.375$^{\circ}$, and 0.5$^{\circ}$, and a ring model with inner and outer 
radii of 0.45$^{\circ}$ and 0.65$^{\circ}$, respectively. We found that the significance of the observed signal exceeds 4$\sigma$ for a point 
source model and for the three first disk models. The highest evaluated significance is 4.7$\sigma$ and is provided by the model of gamma-ray 
emission from a disk of 0.25$^{\circ}$ radius. The inclusion of the three low energy bands (1924-3333 MeV, 3333-5773 MeV, and 5773-10000 MeV) 
to the analysis increases the significance of detection of a tentative gamma-ray excess in the direction of galaxy clusters from 4.7$\sigma$ to 
5.1$\sigma$ for the best-fit model. Fitting a disk model to the data improves the likelihood in comparison to the point-source hypothesis,
but do not provide strong evidence for spatial extension of the observed gamma-ray signal towards the clusters.  
We found no evidence for gamma-ray emission from the structure formation shocks described by the ring model. 

We performed an analysis of a sample of cool-core galaxy clusters and of a sample of non-cool core clusters to study gamma-ray
emission of these two distinct types of galaxy clusters. To divide the 55 selected galaxy clusters into the two samples, we applied the
selection criterion based on the estimated ratio of $\dot{M}/M_{500}$ to our sample of galaxy clusters.  The performed analysis shows
that the contribution of cool-core clusters to the gamma-ray emission from the 55 selected clusters is dominant over that of non-cool
core clusters and that the gamma-ray properties of the cool-core clusteres significantly differs from those of the noncooling
core clusters.

To reveal the origin of the observed gamma-ray excess in the direction of the 55 selected galaxy clusters, we studied independently the signals 
from each of these galaxy clusters. We found that there are 10 galaxy clusters in our sample such that the probability to obtain the observed 
number of photons from each of these clusters within the circular region with a radius of 0.3$^{\circ}$ by chance is less than 5\%. 
We searched for time variability of the signals from these 10 galaxy clusters using the the nearest-neighbor method for testing homogeneity 
of incoming photons in time. The analysis of time variability showed that the homogeneity hypothesis is not valid 
and that a fraction of the observed photons has been emitted by variable gamma-ray sources, such as active galactic nuclei. 

We performed the search for gamma-ray candidate sources from the CGRaBS catalogue within a circle of 0.6$^{\circ}$ ($\approx$2 PSF) 
towards the selected 55 galaxy clusters and found that the gamma-ray candidate sources are present towards eight galaxy clusters.
Six of these eight CGRaBS gamma-ray candidate sources are distant blazars with redshifts larger than 0.35.
We estimated that the significance of a gamma-ray signal after withdrawal of eight galaxy clusters possibly associated with 
CGRaBS candidate sources. We found that 
the significance of gamma-ray detection for this subsample is less than that derived from the sample of 55 galaxy clusters and that
the highest significance level (3.6$\sigma$) is also provided by the disk model with a radius of 0.25$^{\circ}$.  
Comparing the observed fluxes with those expected in the neutral pion decay model, we found that the best-fit value of 
the relativistic-hadron-to-thermal energy density ratio equals $\approx$1.5\% for the spectral index of the parent CR hadron distribution, 
$f(p)\propto p^{-s}$, equal s=4.1.
Given that AGNs are likely contributing to the observed flux this value can be treated as an upper limit on the 
relativistic-hadron-to-thermal energy density ratio, provided that relativistic and thermal components are mixed.

\begin{appendix}

\section{}

\begin{table}
\centering
\caption{The list of the masked point sources}
\begin{tabular}{|c|c|c|}
\hline
2GL source name & $\sqrt{TS}$ at E$>$10 GeV&cluster name\\ 
\hline
2FGLJ0009.9-3206 & 1.7 & A4059\\
2FGLJ0050.6-0929 & 7.7 & A0085\\ 
2FGLJ0051.0-0648 & 2.5 & A0085\\
2FGLJ0055.0-2454 & 1.4 & A0133\\ 
2FGLJ0056.8-2111 & 3.5 & A0133\\ 
2FGLJ0059.2-0151 & 4.7 & A0119\\ 
2FGLJ0118.8-2142 & 4.0 & A0133\\ 
2FGLJ0112.8+3208 & 3.5 & NGC507\\ 2
2FGLJ0122.6+3425 & 5.8 & NGC507\\ 
2FGLJ0309.1+1027 & 4.6 & A0399, A0401\\
2FGLJ0325.1-5635 & 2.1 & A3158\\
2FGLJ0335.3-4501 & 5.7 & A3112\\
2FGLJ0338.2+1306 & 5.0 & 2A0335, ZwIII 54\\
2FGLJ0414.9-0855 & 3.8 & EXO0422\\
2FGLJ0416.8+0105 & 3.4 & NGC1550\\
2FGLJ0423.2-0120 & 9.0 & NGC1550\\
2FGLJ0424.7+0034 & 3.4 & NGC1550\\
2FGLJ0647.7-5132 & 4.0 & A3391\\
2FGLJ0710.5+5908 & 10.1 & A0576\\
2FGLJ0742.6+5442 & --- & A0576\\
2FGLJ0856.6-1105 & 6.0 & A0754\\
2FGLJ0906.2-0906 & 3.1 & A0754\\
2FGLJ1141.9+1550 & 4.2 & A1367\\
2FGLJ1154.0-0010 & 7.1 & MKW4\\
2FGLJ1229.1+0202 & 4.7 & ZwCl1215\\
2FGLJ1256.1-0547 & 16.0 & A1650, A1651\\
2FGLJ1310.9+0036 & 5.5 & A1650\\ 
2FGLJ1313.0-0425 & --- & A1651\\
2FGLJ1315.9-3339 & 4.0 & A3558\\
2FGLJ1345.8-3356 & 1.8 & A3562\\
2FGLJ1347.0-2956 & 4.2 & A3562, A3571\\
2FGLJ1351.1-2749 & 1.3 & A3581\\
2FGLJ1406.2-2510 & 5.1 & A3581\\
2FGLJ1416.3-2415 & 4.5 & A3581\\
2FGLJ1440.9+0611 & 6.1 & MKW8\\
2FGLJ1505.1+0324 & --- & A2029\\
2FGLJ1506.6+0806 & 3.2 & A2029, A2059\\
2FGLJ1506.6+0806 & 3.2 & MKW35, A2063\\
2FGLJ1508.5+2709 & 3.8 & A2065\\
2FGLJ1512.2+0201 & 6.0 & A2029\\
2FGLJ1522.1+3144 & 12.8 & A2065\\
2FGLJ1539.5+2747 & 2.4 & A2065\\
2FGLJ1548.3+1453 & 4.8 & A2147\\
2FGLJ1553.5+1255 & 3.5 & A2147\\
2FGLJ1607.0+1552 & 1.8 & A2147\\
2FGLJ1635.2+3810 & 5.6 & A2199\\
2FGLJ1640.7+3945 & 2.8 & A2199\\
2FGLJ1642.9+3949 & 2.2 & A2199\\
2FGLJ1800.5+7829 & 11.5 & A2256\\
2FGLJ1955.0-5639 & 4.0 & A3667\\
2FGLJ2317.3-4534 & 2.7 & S1101\\
2FGLJ2319.1-4208 & 3.7 & S1101\\
2FGLJ2324.7-4042 & 8.5 & S1101\\
2FGLJ2327.9-4037 & 0.5 & S1101\\
2FGLJ2347.2+0707 & 4.1 & A2657\\
2FGLJ2350.2-3002 & 3.1 & A4038\\
2FGLJ2353.5-3034 & 4.4 & A4038\\
2FGLJ2359.0-3037 & 4.4 & A4038\\  
\hline
\end{tabular}
\label{Table1}
\end{table}

\begin{table}
\centering
\caption{List of photon events within a circular region of 0.3$^{\circ}$ radius
centered on A0119, A3112, A4038, A0400, 2A0335, A1367, MKW4, A2255, A2063, and A3581}
\begin{tabular}{|c|c|c|}
\hline
Cluster name & Arrival time (MET), s & Energy, GeV \\ 
\hline
A0119 & 261033225 & 10.8 \\
A0119 & 267864212 & 28.3 \\ 
A0119 & 290446871 & 150.0 \\
A3112 & 247100190 & 66.7 \\
A3112 & 251029235 & 90 \\
A3112 & 268175993 & 11.2 \\
A4038 & 282124076 & 11.9 \\
A4038 & 310302979 & 20.4 \\
A4038 & 369147989 & 13.1 \\
A0400 & 250037714 & 24.1 \\ 
A0400 & 291576539 & 11.9 \\ 
A0400 & 297933185 & 50.8 \\
A0400 & 330823308 & 10.7 \\
2A0335 & 283348255 & 11.7 \\
2A0335 & 333270630 & 10.8 \\
2A0335 & 352741781 & 36.8 \\
2A0335 & 358456858 & 10.3 \\
A1367 & 260628741 & 48.6 \\
A1367 & 314405927 & 14.9 \\
A1367 & 349727501 & 70 \\
A1367 & 351437188 & 22.3 \\
MKW4 & 241698944 & 14.3 \\
MKW4 & 255282333 & 10.3 \\
MKW4 & 263466545 & 10.9 \\
MKW4 & 323071228 & 11.5 \\
A2255 & 327382581 & 60 \\
A2255 & 338658407 & 12.3 \\
A2255 & 361107399 & 14.6 \\
A2255 & 373040091 & 177.5 \\
A2063 & 245663292 & 18.9 \\
A2063 & 249511419 & 18.0 \\
A2063 & 250975083 & 16.9 \\
A2063 & 333842971 & 12.8 \\
A2063 & 348468092 & 14.1 \\
A3581 & 242221372 & 16.8 \\
A3581 & 252236390 & 11.7 \\
A3581 & 272025793 & 15.8 \\
A3581 & 292681335 & 12.2 \\
A3581 & 300957764 & 11.9 \\
\hline
\end{tabular}
\label{Table4}
\end{table} 

\end{appendix}

\bibliographystyle{aa}
\bibliography{ref}

\end{document}